\documentclass[showpacs,preprint,amsmath,amssymb,pre]{revtex4-1}

\usepackage{amsmath,amssymb}
\usepackage{graphicx}

\begin{document}

\title{Aging transition in systems of oscillators with global distributed-delay coupling}

\author{B. Rahman}
\author{K.B Blyuss}
\author{Y.N. Kyrychko}
\email{y.kyrychko@sussex.ac.uk}
\affiliation{Department of Mathematics, University of Sussex, Falmer, Brighton, BN1 9QH, United Kingdom}

\date{\today}

\begin{abstract} 
We consider a globally coupled network of active (oscillatory) and inactive (non-oscillatory) oscillators with distributed-delay coupling. Conditions for aging transition, associated with suppression of oscillations, are derived for uniform and gamma delay distributions in terms of coupling parameters and the proportion of inactive oscillators. The results suggest that for the uniform distribution increasing the width of distribution for the same mean delay allows aging transition to happen for a smaller coupling strength and a smaller proportion of inactive elements. For gamma distribution with sufficiently large mean time delay, it may be possible to achieve aging transition for an arbitrary proportion of inactive oscillators, as long as the coupling strength lies in a certain range.
\end{abstract}


\maketitle 

\section{Introduction}

In the studies of large complex networks, an important practical consideration is the robustness of network structure and function under external perturbations and malfunctions. Of particular interest is the ability of large systems of coupled units to continue maintaining their dynamical activity when some part of the system becomes inactive due to deterioration or inactivation. In the last two decades this question has featured prominently in a number of different contexts \cite{sasai15}, ranging from power grids \cite{rohden12} to cardiac pacemaker cells \cite{glass01,miake02}, circadian clocks \cite{aton2005,bernard07,fukuda07}, and arrays of Josephson junctions \cite{wies1996}. There exist various scenarios how coupling between elements of the network can result in the suppression of otherwise stable endogenous oscillations. {\it Amplitude death} (AD), which refers to coupling-induced suppression of oscillations, has been extensively studied in systems of coupled oscillators with both instantaneous \cite{aron90,mirollo1990}, and time-delayed coupling \cite{ramana1998,reddy1999,ramana2000,atay2003}. While in the case of instantaneous coupling AD can only occur provided the frequencies of oscillators are sufficiently different, when there is a delay in the coupling, AD can happen even for identical oscillators. Furthermore, recent results have shown that when the time delay itself is not just a single discrete delay but rather some delay distribution, this can also have a profound effect on amplitude death and the emergence of different kinds of phase-locked solutions \cite{kyrychko2011, kyrychko2013,kyrychko2014}. Whilst AD represents quenching of oscillations through stabilisation of a previously unstable fixed point, another possibility to suppress oscillations is through {\it oscillation death}, which means the emergence of a stable coupling-induced inhomogeneous steady state that did not exist in the original system \cite{kos09,kos10,zak13}.

Quite often, the system under consideration can be viewed as a collection of coupled active and inactive oscillators. Such systems can exhibit an interesting phenomenon called {\it aging transition} (AT), whereby as the fraction of inactive oscillators increases, at some point the system may completely lose its oscillatory dynamics, effectively exhibiting an amplitude death \cite{daido2004,zou2015,huang14}. Aging transition was originally discovered in globally coupled Stuart-Landau oscillators \cite{daido2004}, and has been subsequently analysed in a variety of networks with different types of coupling, including locally coupled networks \cite{daido08,daido11}, globally coupled networks \cite{daido2006,pazo2006,daido13,yao14,zou2013}, multilayered \cite{morino2011} and complex networks \cite{sasai15,he2013,huang14,tanaka12,tanaka14}. Several authors have also studied systems of oscillators with different periods \cite{tanaka2010}, as well as mixed populations of passive, oscillatory and excitable cells \cite{kryukov2008}. Thakur {\it et al.} \cite{thakur2014} have investigated the dynamics of $N$ globally coupled active and inactive oscillators with discrete time delay in the coupling, and they found that time delay can be used to decrease the coupling strength and the proportion of inactive oscillators above which the system exhibits aging transition.

In many realistic contexts, time delays in the interactions between oscillators are themselves not constant but rather satisfy a certain distribution, and it is important to correctly account for this in the mathematical models. One prominent example of such a situation is the dynamics of a population of single-cell circadian oscillators in the suprachiasmatic nucleus (SCN) of mammalian brain \cite{liu97,rep02,indic07}. Whilst the genetic regulation of individual SCN neurons can be described quite well using models with discrete time delays \cite{smol01}, interactions between neurons may involve parallel pathways with different axon sizes and lengths, which results in different distributions of signals between neurons \cite{smol99,atay06,hutt13}. Hence, to understand the collective behaviour of multiple SCN neurons it is essential to include distributed time delays in the connections between them \cite{smol99}, which can shed light on possible suppression of oscillatory activity due to environmental effects \cite{webb09} or age-related deterioration \cite{aujard01}. Similarly, when one considers a plant circadian system, there are two major types of connections between plant cells: local connections via plasmodesmata, and long-range connections via vascular bundles \cite{ding98,lude02}. Since most of the plant cells are able to exhibit self-sustained circadian oscillations, whereas the cells in the vascular bundles lack the necessary genetic machinery for generating such oscillations \cite{lude02}, plants give another example of a system of coupled active and inactive oscillators that can exhibit aging transition. Furthermore, experimental observations of the spatiotemporal activity in plant leaves strongly suggest that the interactions between different groups of plant cells should be described using delay distributions \cite{rasch01,fukuda07}. Another practical example where distributed delays arise naturally in the coupling between active and inactive oscillators is the coarse-scale models of power grids \cite{fila08,rohden12,molina11,hart12}. In such models, many of which are based on large systems of Kuramoto-like oscillators, coupling can result in the suppression of synchronised oscillatory behaviour \cite{dorf13,menck14} that can manifest itself in the form of cascading power grid failures.

Motivated by the above-mentioned examples, in this paper we analyse an aging transition in a system of $N$ globally coupled Stuart-Landau oscillators with distributed-delay coupling
\begin{equation}\label{sys1}
\begin{array}{l}
\dot{Z}_j(t)={\displaystyle(\alpha_j+i\omega)Z_j(t)-|Z_j(t)|^2Z_j(t)+\frac{k}{N}\sum_{i=1}^{N}\left[\int_0^{\infty}g(t')Z_i(t-t')dt'
	-Z_j(t)\right]}\\\\
-{\displaystyle\frac{k}{N}\left[\int_0^{\infty}g(t')Z_j(t-t')dt'-Z_j(t)\right],\qquad j=1,...,N,}  
\end{array}
\end{equation}
where $Z_j\in\mathbb{C}$, $\omega$ is the natural frequency of oscillations, $\alpha_j$ are bifurcation parameters that control whether in the absence of coupling the oscillator $j$ exhibits a stable periodic solution $(\alpha_j>0)$ or converges to a stable trivial equilibrium $(\alpha_j<0)$, $k$ is the coupling strength, and the last  term in the right-hand side of (\ref{sys1}) removes self-coupling from the system. The kernel $g$ is taken to be positive-definite and normalised to unity, i.e.
\[
g(u) \geq 0,\quad {\displaystyle \int_0^{\infty}g(u)du=1}.
\]
If the distribution kernel is taken in the form of the Dirac delta function $g(u)=\delta(u)$, one obtains a system of globally instantaneously coupled oscillators that has been extensively studied \cite{mirollo1990,aizawa1976, yamaguchi1984, shiino1989, ermentrout1990}, while for $g(u)=\delta(u-\tau)$, the model (\ref{sys1}) reduces to a system with a single discrete time delay studied in \cite{zou2013,thakur2014}. By focusing on the dynamics of system (\ref{sys1}), we extend the work of Thakur {\it et al.} \cite{thakur2014} in the direction of a more general and more realistic case of distributed delay. In order to gain a better understanding of the system behaviour inside the stability regions, we will numerically compute characteristic eigenvalues. By considering uniform and then gamma-distributed delay kernel we will show that it is not only the mean delay and the width of the distribution, but also the actual shape of the distribution that affects the amplitude death in systems with distributed-delay coupling.

Following Daido and Nakanishi \cite{daido2004}, let $p$ denote the fraction of inactive oscillators with the bifurcation parameter $\alpha_j=-b$ $(b>0)$, and $q=1-p$ be the fraction of active oscillators with the bifurcation parameter $\alpha_j=a$ $(a>0)$. To analyse the aging transition in the system (\ref{sys1}) we use a mean-field approximation \cite{daido2004,thakur2014} which assumes that all active and inactive oscillators achieve synchronization in their respective subpopulations (see \cite{huang14} for further numerical evidence in support of this), i.e. we write $Z_j(t)=A(t)$ for all active elements and $Z_j(t)=I(t)$ for all inactive oscillators. This reduces the original high-dimensional system (\ref{sys1}) to the following
\begin{equation}\label{sys2}
\begin{array}{llll}
\dot{A}(t)=&{\displaystyle\left[a-k\left(1-\frac{1}{N}\right)+i\omega-|A(t)|^2\right]A(t)} \\\\
&+{\displaystyle k\left(q-\frac{1}{N}\right)\int_0^{\infty} g(t')A(t-t')dt'+kp\int_0^{\infty}g(t')I(t-t')dt',} \\ \\  
\dot{I}(t)=&{\displaystyle\left[-b-k\left(1-\frac{1}{N}\right)+i\omega-|I(t)|^2\right]I(t)}\\ \\
&+{\displaystyle k\left(p-\frac{1}{N}\right)\int_0^{\infty} g(t')I(t-t')dt'+kq\int_0^{\infty}g(t')A(t-t')dt'}. 
\end{array}
\end{equation}
Linearisation of this system near the trivial steady state $(A,I)=(0,0)$ yields the following characteristic equation for eigenvalues $\lambda$
\begin{equation}\label{eq0}
\begin{array}{l}
\displaystyle{\left[a-k\left(1-\frac{1}{N}\right)+i\omega+k\left(q-\frac{1}{N}\right)\widehat{G}(\lambda)-\lambda\right]}\\\\
\displaystyle{\times\left[-b-k\left(1-\frac{1}{N}\right)+i\omega+k\left(p-\frac{1}{N}\right)\widehat{G}(\lambda)-\lambda\right]-k^2pq\widehat{G}^2(\lambda)=0,}
\end{array}
\end{equation}
where
\begin{center}
	${\displaystyle\widehat{G}(\lambda)=\int_0^{\infty}e^{-\lambda u}g(u)du,}$ 
\end{center}
is the Laplace transform of the function $g(u)$.

The rest of the paper is organised as follows. In Section \ref{sec2} aging transition in system (\ref{sys2}) is analysed for the case of uniformly-distributed delay kernel. This includes finding analytical expressions for stability boundaries of the trivial steady state, as well as numerical computation of the eigenvalues of the corresponding characteristic equations. Section \ref{sec3} is devoted to the analysis of aging transition for the case of gamma-distributed delay kernel. We illustrate how regions of amplitude death are affected by the coupling parameters and characteristics of the delay distribution. The paper concludes with a summary and discussion in Section \ref{sec4}.
 
\section{Uniform distribution kernel} \label{sec2}

To make analytical progress with analysis of equation (\ref{eq0}), we begin by considering the {\it uniformly-distributed delay kernel}
\begin{equation}\label{eq1}
g(u)=\left\{\begin{array}{ccc}
\displaystyle{\frac{1}{2\sigma} }& \text{for} & \tau-\sigma \leq u \leq \tau+\sigma, \\ \\
\displaystyle{0} & & \text{otherwise},
\end{array}\right.
\end{equation}
which has the mean time delay $\tau$ and the width $2\sigma$. This type of time delay has been successfully used to study dynamics of ecological models \cite{nun85}, genetic regulation \cite{barrio06}, and to model response time in power grid networks \cite{molina11}. Laplace transform of the uniform distribution (\ref{eq1}) can be readily found as
\begin{equation}\label{eq2}
\displaystyle{\widehat{G}(\lambda)=\frac{1}{2\sigma\lambda}e^{-\lambda\tau}(e^{\lambda\sigma}-e^{-\lambda\sigma})=e^{-\lambda\tau}\frac{\sinh (\lambda\sigma)}{\lambda\sigma}}.
\end{equation} 
Substituting this expression into the characteristic equation (\ref{eq0}) and looking for characteristic roots in the form $\lambda=i\xi$, separating real and imaginary parts gives
\begin{equation}\label{eq3}
\begin{array}{l}
{\displaystyle\left[a-k\left(1-\frac{1}{N}\right)+k\left(q-\frac{1}{N}\right)\cos(\xi\tau)\gamma(\xi,\sigma)\right]\left[-b-k\left(1-\frac{1}{N}\right)+k\left(p-\frac{1}{N}\right)\cos(\xi\tau)\gamma(\xi,\sigma)\right]}\\\\ 
{\displaystyle-\left[\omega-\xi+k\left(q-\frac{1}{N}\right)\sin(\xi\tau)\gamma(\xi,\sigma)\right]\times\left[\omega-\xi+k\left(p-\frac{1}{N}\right)\sin(\xi\tau)\gamma(\xi,\sigma)\right]} \\\\ 
{\displaystyle=k^2pq\cos(2\xi\tau)\gamma^2(\xi,\sigma),}\\\\
{\displaystyle\quad\left[-b-k\left(1-\frac{1}{N}\right)+k\left(p-\frac{1}{N}\right)\cos(\xi\tau)\gamma(\xi,\sigma)\right]\left[\omega-\xi+k\left(q-\frac{1}{N}\right)\sin(\xi\tau)\gamma(\xi,\sigma)\right]}\\\\
{\displaystyle+\left[a-k\left(1-\frac{1}{N}\right)+k\left(q-\frac{1}{N}\right)\cos(\xi\tau)\gamma(\xi,\sigma)\right]\times\left[\omega-\xi+k\left(p-\frac{1}{N}\right)\sin(\xi\tau)\gamma(\xi,\sigma)\right]}\\\\
{\displaystyle=-k^2pq\sin(2\xi\tau)\gamma^2(\xi,\sigma),}
\end{array}
\end{equation}
where 
\[
\gamma(\xi,\sigma)=\frac{\sin(\xi\sigma)}{\xi\sigma}.
\]
Solution of the system of equations (\ref{eq3}) gives an implicit expression for the boundary of AT, i.e. the suppression of oscillations, in terms of system parameters. This boundary can be described as $\max\{{\rm Re}(\lambda)\}=0$, so that $\max\{{\rm Re}(\lambda)\}\leq 0$ inside the amplitude death regions.

In order to better understand what is actually happening inside the corresponding stability regions, we use a pseudospectral method described in \cite{breda2006} and implemented in the traceDDE toolbox for Matlab to numerically compute the characteristic eigenvalues of equation (\ref{eq0}). To this end, we introduce auxiliary real variables $A=A_r+iA_i$ and $I=I_r+iI_i$, and rewrite the linearised system with uniformly-distributed delay kernel (\ref{eq1}) as follows
\begin{equation}\label{sys3}
\dot{\bf{z}}(t)=L_0{\bf z}(t)+\frac{1}{2\sigma}\int_{-(\tau+\sigma)}^{-(\tau-\sigma)}M{\bf{z}}(t+s)d\tau,
\end{equation}
where ${\bf z}=(A_r,A_i,I_r,I_i)^T$,
\begin{figure}
\hspace{-0.7cm}
	\centering
	\includegraphics[scale=0.16]{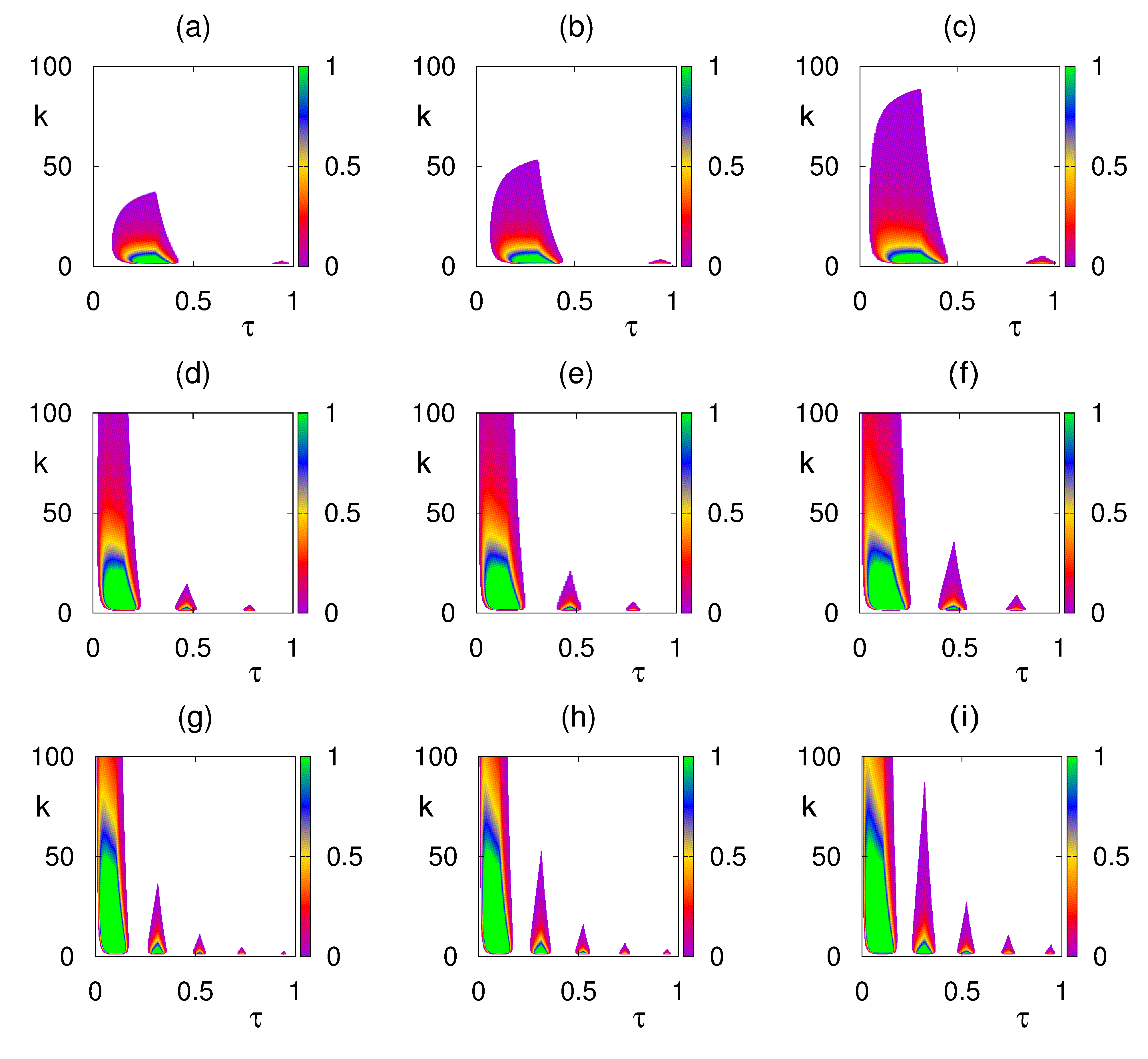}\vspace{-0.5cm}
	\caption{Regions of AD (aging transition) for the system (\ref{sys2}) with uniform distribution kernel for 
		$a=2$, $b=1$, $N=500$, $\sigma=0$, (a)-(c) $\omega=10$, (d)-(f) $\omega=20$, (g)-(i) $\omega=30$. Colour code denotes $[-\max\{{\rm Re}(\lambda)\}]$ for $\max\{{\rm Re}(\lambda)\}\leq 0$. (a), (d), (g) $p=0.3$. (b), (e), (h) $p=0.4$. (c), (f), (i) $p=0.5$.}\vspace{-0.3cm}
	\label{fig1}
\end{figure}
\[
L_0=\left (
\begin{array}{cccc}
a-k\left(1-\frac{1}{N}\right) & -\omega & 0 & 0 \\
\omega & a-k\left(1-\frac{1}{N}\right) & 0 & 0 \\
0 & 0 & -b-k\left(1-\frac{1}{N}\right) &  -\omega \\
0 & 0 & \omega & -b-k\left(1-\frac{1}{N}\right) 
\end{array}\right), \]
and
\[
M=\left( 
\begin{array}{cccc}
k\left(q-\frac{1}{N}\right) & 0 & kp & 0  \\
0 & k\left(q-\frac{1}{N}\right) & 0 & kp \\
kq & 0 & k\left(p-\frac{1}{N}\right) & 0 \\
0 & kq & 0 & k\left(1-\frac{1}{N}\right) 
\end{array}\right).
\]
 
\begin{figure}
\hspace{-0.7cm}
	\centering
	\includegraphics[scale=0.16]{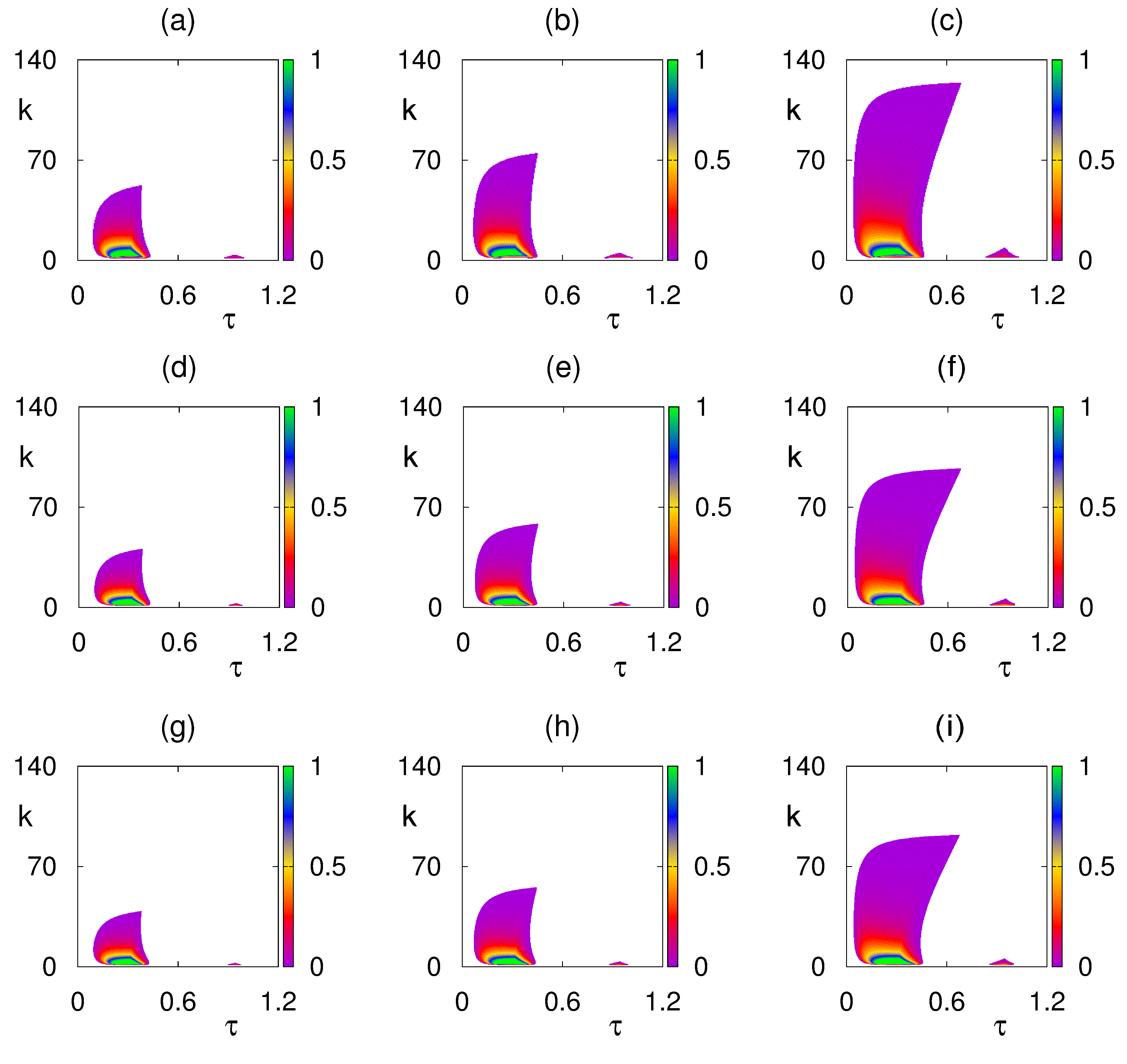}\vspace{-0.5cm}
	\caption{Regions of AT (aging transition) for the system (\ref{sys2}) with uniform distribution kernel for 
		$a=2$, $b=1$, $\sigma=0.02$, $\omega=10$, (a)-(c) $N=4$, (d)-(f) $N=20$, (g)-(i) $N=500$. Colour code denotes $[-\max\{{\rm Re}(\lambda)\}]$ for $\max\{{\rm Re}(\lambda)\}\leq 0$. (a), (d), (g) $p=0.3$. (b), (e), (h) $p=0.4$. (c), (f), (i) $p=0.5$.
	}\vspace{-0.3cm}
	\label{fig2}
\end{figure}

When $\sigma=0$, the last term in the equation (\ref{sys3}) turns into $M{\bf{z}}(t-\tau)$, and system (\ref{sys2}) reduces to the system with a single discrete time delay $\tau$ analysed in \cite{thakur2014}.  Figure~\ref{fig1} shows the regions of amplitude death together with the magnitude of the real part of the leading eigenvalue of the characteristic equation (\ref{eq3}) in the $(\tau,k)$ parameter space for the case of a single discrete delay, i.e. $\sigma=0$. One observes that increasing the natural frequency of oscillations $\omega$ leads to the increase in the number of stability islands and their size. Whilst Thakur et al. \cite{thakur2014} have also noticed the increasing size of stability islands associated with increasing $p$, we also note the appearance of additional stability islands for higher values of time delay $\tau$ or for higher natural frequency $\omega$.

Before looking into the effects of the width of distribution $\sigma$, it is important to understand the role of the system size, i.e. how the overall number of oscillators $N$ influences the AT. Figure~\ref{fig2} shows that for the same $a$, $b$ and the width of delay distribution $\sigma$, increasing the number of oscillators $N$ leads to shrinking of the stability regions. However, this only occurs for relatively small numbers of oscillators, at which point stability regions reach some stationary configuration, beyond which they appear to be unaffected by subsequent increases of $N$ from $20$ to $500$.

\begin{figure}
\hspace{-0.7cm}
	\centering
	\includegraphics[scale=0.16]{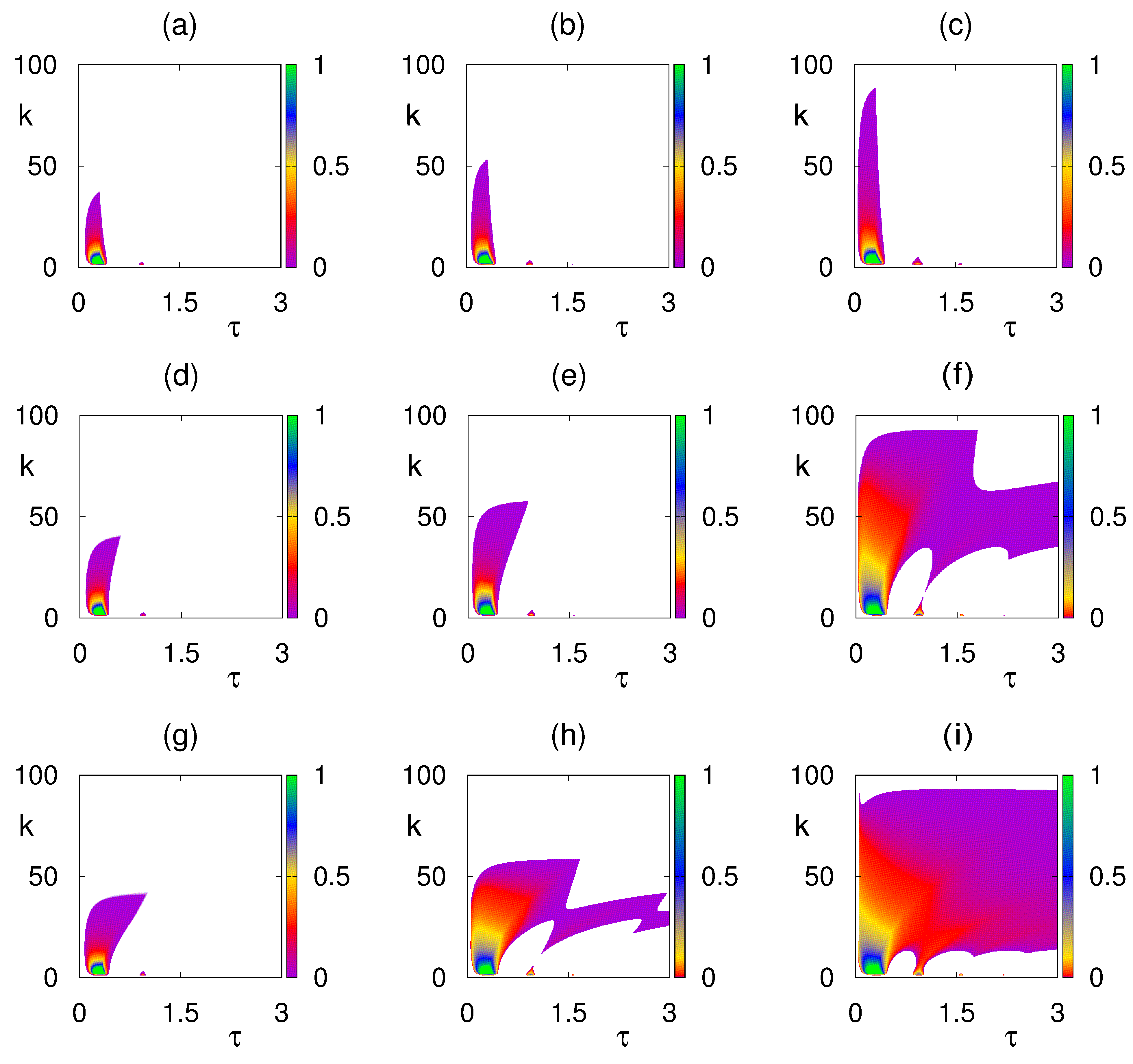}\vspace{-0.5cm}
	\caption{Regions of AT (aging transition) for the system (\ref{sys2}) with uniform distribution kernel for 
		$a=2$, $b=1$, $N=500$, $\omega=10$, (a)-(c) $\sigma=0$, (d)-(f) $\sigma=0.03$, (g)-(i) $\sigma=0.06$. Colour code denotes $[-\max\{{\rm Re}(\lambda)\}]$ for $\max\{{\rm Re}(\lambda)\}\leq 0$. (a), (d), (g) $p=0.3$. (b), (e), (h) $p=0.4$. (c), (f), (i) $p=0.5$.
	}\vspace{-0.3cm}
	\label{fig3}
\end{figure}

To investigate the role of the width of uniform delay distribution $\sigma$, we now compare the results for the discrete and the uniformly-distributed delays with the same mean time delay $\tau$. Figure~\ref{fig3} shows that as $\sigma$ grows, this leads to an increase in the size of AD parameter regions, and the stability islands grow dramatically until they merge into a single continuous region along $\tau$ axis for higher proportions of the inactive oscillators $p$. One should note that while the difference between the delta- and uniformly-distributed delays is small for sufficiently small $p$, it becomes much more pronounced as $p$ increases.
Already in the case when exactly half of all oscillators are inactive $(p=0.5)$, whereas for $\sigma=0$ there were still just three isolated death islands, for $\sigma=0.03$ there is a whole range of coupling strengths where amplitude death can be achieved for an arbitrary value of the mean time delay $\tau$, provided it is greater than some minimum value. One can observe an interesting interplay between $\sigma$ and $p$ which controls whether and when this transition to unbounded AD regions occurs. If the number of inactive oscillators $p$ is very small, then although increasing the width of the delay distribution $\sigma$ does increase the size of the stability regions, it may not be enough to result in the unbounded AD region. As $p$ increases, it becomes possible to achieve unbounded AD regions, and the required value of $\sigma$ is itself decreasing with $p$.

\begin{figure}
	\centering
	\includegraphics[scale=0.9]{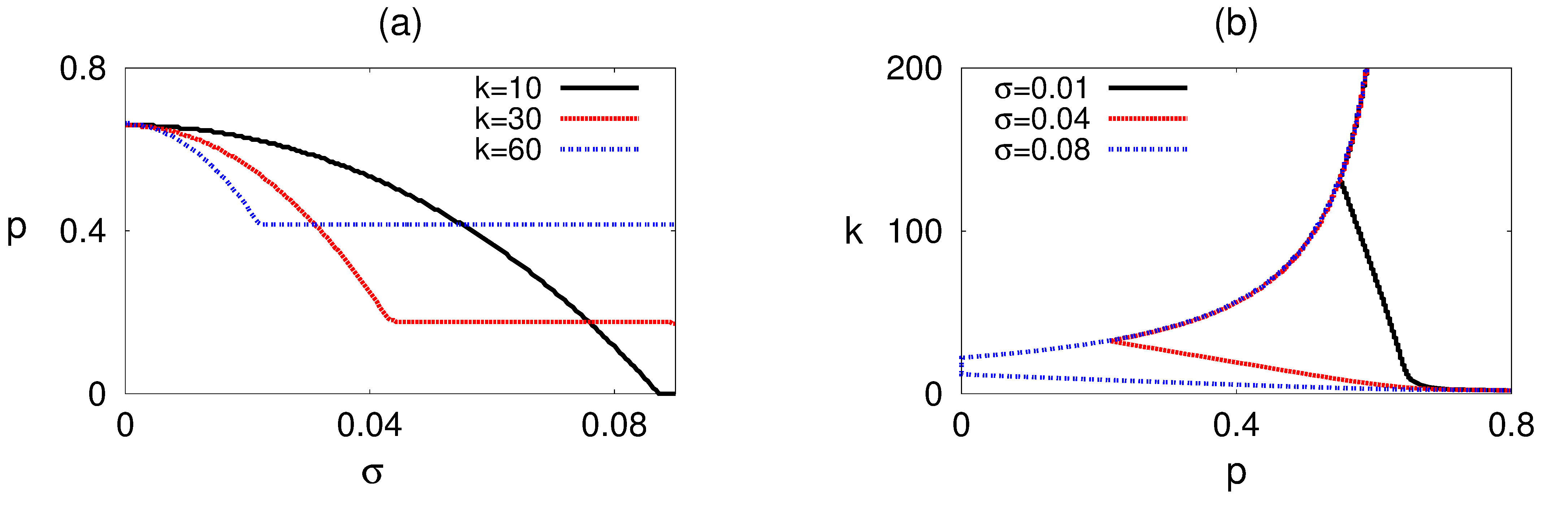}
	\caption{Boundaries of AT for the system (\ref{sys2}) with uniform distribution kernel for $a=2$, $b=1$, $\omega=10$, $\tau=0.5$, $N=500$. Suppression of oscillations occurs below the curves in (a), and to the right of the boundary curves in (b).}
	\label{twoboundary}
\end{figure}

Figure~\ref{twoboundary} illustrates how the AT boundary of the trivial steady state depends on the parameters $\sigma$, $p$, and $k$.
It is noteworthy that as the coupling strength increases, the size of AD regions in the $\sigma$-$p$ plane decreases, and, in fact, AD happens for a smaller range of distribution widths for the same fraction of inactive oscillators $p$. Figure~\ref{twoboundary}(b) suggests that as $\sigma$ increases, AT occurs for smaller values of $p$ for the same coupling strength $k$, thus implying that increasing the width of distribution for the same mean time delay can make the aging transition take place sooner that it would happen in the case of a discrete time delay.

\section{Gamma distribution kernel}
\label{sec3}

As our next example we consider the {\it gamma distribution}, which can be written as follows
\begin{equation}\label{eqg1}
\displaystyle{g(u)=\frac{u^{r-1}\zeta^{r} e^{-\zeta u}}{(r-1)!}},
\end{equation}
where $\zeta,r\geq 0$, and $r$ is an integer. For $r=1$ this is an exponential distribution, also called a \emph{weak delay kernel}, and for $r=2$, it is known as the {\it strong delay kernel}. Gamma delay distribution has been extensively studied in models of population dynamics \cite{cooke82,blythe85}, intracellular dynamics of HIV infection \cite{mit98}, machine tool vibrations \cite{step88}, and power grid stability \cite{zhu11,chen14}. The distribution (\ref{eqg1}) has the mean delay
\begin{equation}\label{taum}
\tau_m=\int_0^{\infty}ug(u)du=\frac{r}{\zeta},
\end{equation}
and the variance
\begin{equation}
\sigma^2=\int_0^{\infty}(u-\tau_m)^2g(u)du=\frac{r}{\zeta^2}.
\end{equation}
To analyse stability of the trivial steady state of the system (\ref{sys2}) with the gamma delay distribution kernel (\ref{eqg1}), one can either use the characteristic equation (\ref{eq0}) with the Laplace transform of the distribution kernel given by
\begin{equation}\label{eqg2}
\widehat{G}(\lambda)=\left(\frac{\zeta}{\lambda+\zeta}\right)^r,
\end{equation}
or use the so-called \emph{linear chain trick} \cite{MacD} that allows one to replace the original system with distributed delays by a system of $(r+1)$ ordinary differential equations. Considering the gamma distributed kernel in (\ref{sys2}) with $r=1$ and introducing new variables
\[
\begin{array}{ll}
X_A(t)&={\displaystyle\int_0^{\infty} \zeta e^{-\zeta s}A(t-s)ds,} \\ \\
X_I(t)&={\displaystyle\int_0^{\infty} \zeta e^{-\zeta s}I(t-s)ds},
\end{array}
\]
one can rewrite the system (\ref{sys2}) as follows
\begin{equation}\label{sysg1}
\left\{ \begin{array}{ll}
\dot{A}(t)&={\displaystyle\left[a-k\left(1-\frac{1}{N}\right)+i\omega\right]A(t)+k\left(q-\frac{1}{N}\right)X_A(t)+kpX_I(t),} \\ \\  
\dot{I}(t)&={\displaystyle\left[-b-k\left(1-\frac{1}{N}\right)+i\omega\right]I(t)+k\left(p-\frac{1}{N}\right)X_I(t)+kqX_A,} \\ \\
\dot{X}_A(t)&={\displaystyle\zeta A(t)-\zeta X_A(t),} \\ \\
\dot{X}_I(t)&={\displaystyle\zeta I(t)-\zeta X_I(t)}.
\end{array}\right.
\end{equation}
Substituting the Laplace transform (\ref{eqg2}) with $r=1$ into the characteristic equation (\ref{eq0}), or linearising system (\ref{sysg1}) near the trivial steady state $(0,0,0,0)$, gives the following polynomial equation for eigenvalues $\lambda$
\begin{figure}
	\centering
	\includegraphics[scale=0.055]{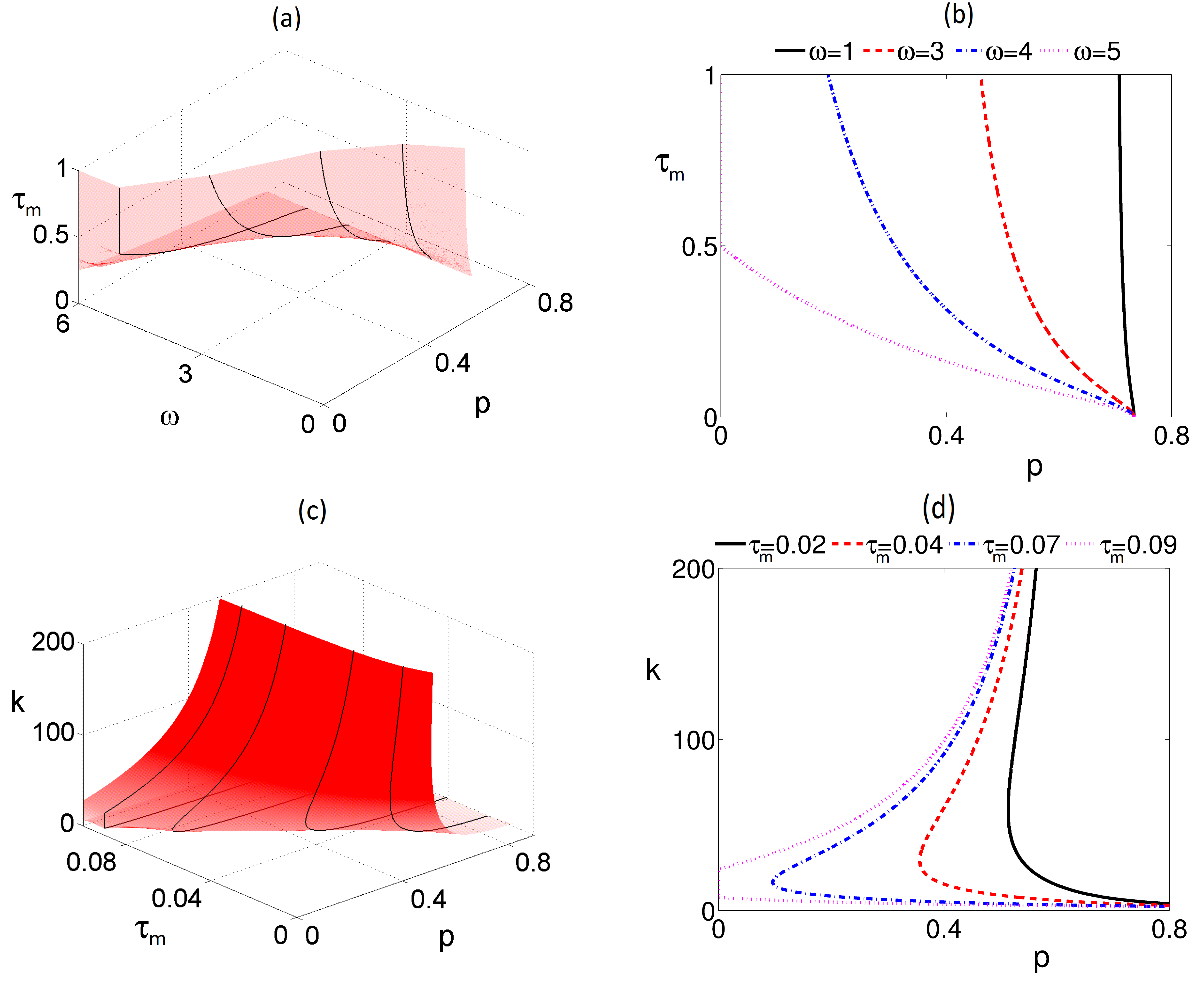}
	\caption{AT boundary of the system (\ref{sysg1}) with the weak delay distribution kernel for $a=2$, $b=1$, $N=500$. (a)-(b) $k=10$, (c)-(d) $\omega=10$. Suppression of oscillations occurs below and to the right of the surface in (a),(c), and to the right of the boundary curves in (b),(d).}
	\label{p_alphaK}
\end{figure}
\begin{equation}\label{eqg3}
\begin{array}{lll}
&{\displaystyle\quad\left[(\lambda+\zeta)\left(a-k\left(1-\frac{1}{N}\right)+i\omega-\lambda\right)+\zeta k\left(q-\frac{1}{N}\right)\right]}\\\\
&{\displaystyle\times\left[(\lambda+\zeta)\left(-b-k\left(1-\frac{1}{N}\right)+i\omega-\lambda\right)+\zeta k\left(p-\frac{1}{N}\right)\right]}\\\\
&-\alpha^2k^2pq=0.
\end{array}
\end{equation}
This equation can be solved numerically to understand the role of different parameters in the aging transition for the case of the gamma distribution kernel. 

Figure~\ref{p_alphaK} illustrates the aging transition boundary in terms of $p$, $\tau_m$, $k$ and $\omega$ for the weak delay distribution (\ref{eqg1}) with $r=1$. For a fixed coupling strength $k$ and sufficiently large natural frequency $\omega$, the AT can occur for an arbitrary fraction of inactive oscillators $p$, as long as the mean time delay $\tau_m$ is sufficiently large. In this case, for smaller values of $\tau_m$ there is a certain minimum $p$ required for AT, and the value of this critical $p$ decreases with $\tau_m$. For smaller values of $\omega$, the AT can only occur if $p$ exceeds a certain threshold that itself decreases with $\tau_m$ until it reaches some value that is independent of $\tau_m$ if $\tau_m$ is sufficiently high. Similarly, for a fixed natural frequency $\omega$ and larger values of $\tau_m$, there is a range of coupling strengths $k$, for which the AT can occur for an arbitrary proportion $p$ of inactive oscillators. For smaller mean time delay $\tau_m$, the AT can only take place for a sufficiently large value of $p$. Interestingly, this critical $p$ initially decreases and then monotonically increases with $k$.

Considering the case of the {\it strong delay kernel} with $r=2$, we use the same strategy as for the weak kernel and introduce new variables as follows
\[
\begin{array}{l}
{\displaystyle X_A(t)=\int_0^{\infty} \zeta e^{-\zeta s}A(t-s)ds,\quad Y_A(t)=\int_0^{\infty} \zeta^2 s e^{-\zeta s}A(t-s)ds,} \\ \\
{\displaystyle X_I(t)=\int_0^{\infty} \zeta e^{-\zeta s}I(t-s)ds,\quad Y_I(t)=\int_0^{\infty} \zeta^2 s e^{-\zeta s}I(t-s)ds.}
\end{array}
\]
This allows one to rewrite the system (\ref{sys2}) in the equivalent form
\begin{equation}\label{sysgs2}
\left\{
\begin{array}{l}
{\displaystyle \dot{A}(t)=\left[a-k\left(1-\frac{1}{N}\right)+i\omega\right]A(t)+k\left(q-\frac{1}{N}\right)Y_A(t)+kpY_I(t),}\\ \\  
{\displaystyle \dot{I}(t)=\left[-b-k\left(1-\frac{1}{N}\right)+i\omega\right]I(t)+k\left(p-\frac{1}{N}\right)Y_I(t)+kqY_A,}\\ \\
\dot{X}_A(t)=\zeta A(t)-\zeta X_A(t), \\ \\
\dot{Y}_A(t)=\zeta X_A(t)-\zeta Y_A(t), \\ \\
\dot{X}_I(t)=\zeta I(t)-\zeta X_I(t), \\ \\
\dot{Y}_I(t)=\zeta X_I(t)-\zeta Y_I(t).
\end{array}\right.
\end{equation}

Linearising this system near the trivial steady state yields the following equation for characteristic eigenvalues $\lambda$
\begin{equation}\label{eqg4}
\begin{array}{l}
{\displaystyle\left[(\lambda+\zeta)^2\left(a-k\left(1-\frac{1}{N}\right)+i\omega-\lambda\right)+\zeta^2 k\left(q-\frac{1}{N}\right)\right]}\\\\
{\displaystyle\times\left[(\lambda+\zeta)^2\left(-b-k\left(1-\frac{1}{N}\right)+i\omega-\lambda\right)+\zeta^2 k\left(p-\frac{1}{N}\right)\right]}\\\\
-\zeta^4k^2pq=0.
\end{array}
\end{equation}

\begin{figure}
	\centering
	\includegraphics[scale=0.055]{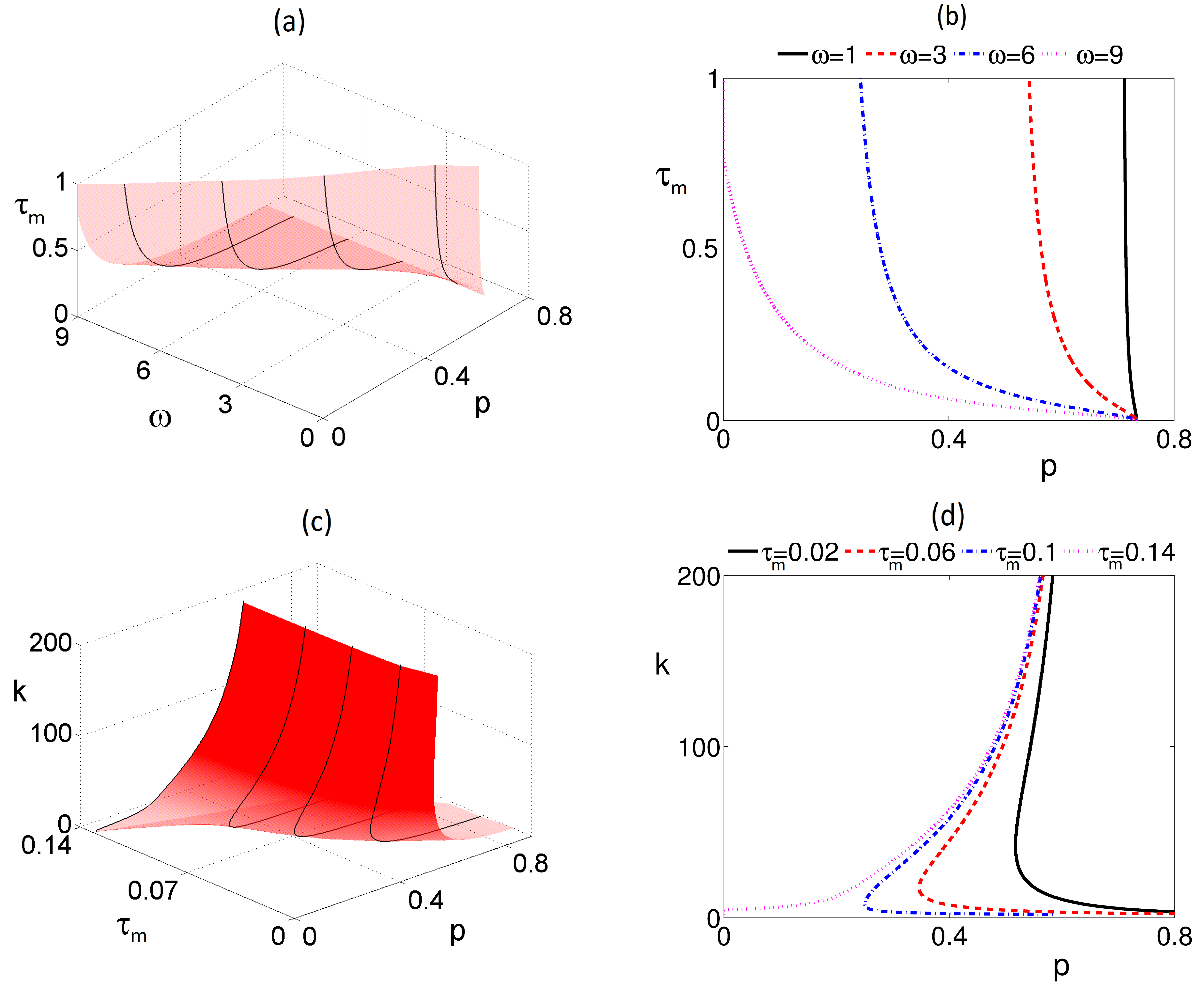}
	\caption{AT boundary of the system (\ref{sysgs2}) with the strong delay distribution kernel for $a=2$, $b=1$, $N=500$. (a)-(b) $k=10$, (c)-(d) $\omega=10$. Suppression of oscillations occurs below and to the right of the surface in (a),(c), and to the right of the boundary curves in (b),(d).}
	\label{strong_p_alphaK}
\end{figure}

Figure~\ref{strong_p_alphaK} illustrates AT boundaries for the strong delay distribution kernel (\ref{eqg1}) with $r=2$. These boundaries exhibit the behaviour qualitatively similar to that for the weak kernel, i.e. for the fixed coupling strength $k$ increasing the natural frequency $\omega$ increases the range of $\tau_m$, for which AT can occur with an arbitrary fraction of inactive oscillators $p$. However, one should note that for the same values of $\omega$ and the mean time delay $\tau_m$, in the case of the strong delay kernel a higher fraction of inactive oscillators $p$ is required to achieve the aging transition.
\begin{figure}
	\hspace{-0.6cm}
		\centering
		\includegraphics[scale=0.1]{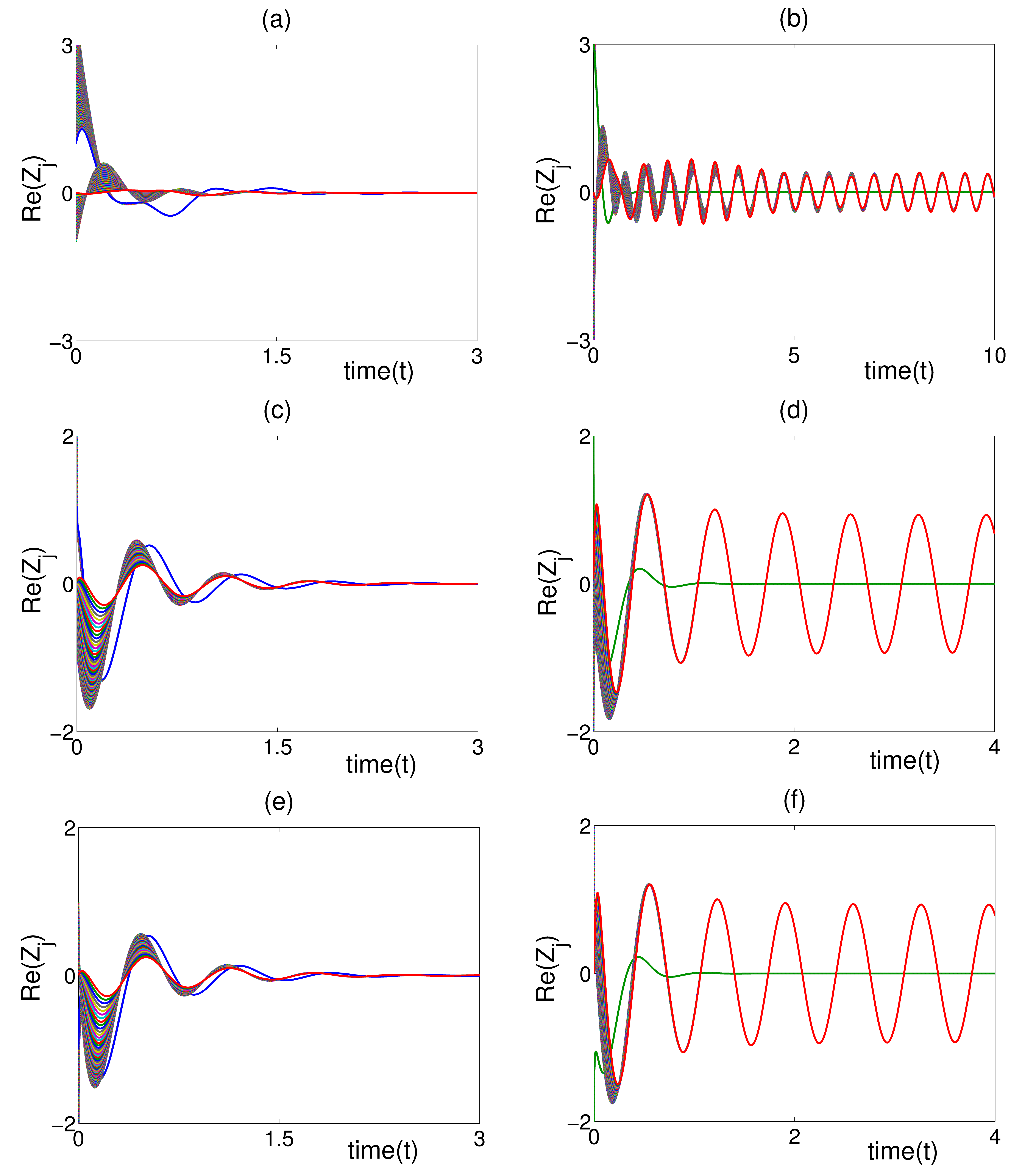}
	\caption{(a)-(b) Solutions of the system (\ref{sys1}) with uniformly-distributed kernel for $\sigma=0.01$, $\tau=0.5$. (c)-(d) Solutions of the system (\ref{sys1}) with weak delay kernel for $\tau_m=0.02$. (e)-(f) Solutions of the system (\ref{sys1}) with strong delay kernel for $\tau_m=0.02$. Other parameters, $a=2$, $b=1$, $\omega=10$, $N=500$, $k=5$. (a),(c),(e) $p=0.8.$ (b),(d),(f) $p=0.2$. Red corresponds to active oscillators, blue/green to inactive oscillators.}
	\label{fig7}
\end{figure}
Furthermore, for sufficiently large $\tau_m$, while for the weak delay kernel AT can already happen for an arbitrary value of $p$, in the case of the strong delay, there may still be a bound on the minimum value of $p$ to achieve aging transition.

To illustrate the dynamics of aging transition for different delay kernels, we show in Fig.~\ref{fig7} the behaviour of system (\ref{sys1}) before and after the aging transition in the case of uniform, weak and strong delay kernels, with the following initial conditions
\[\begin{array}{l}
Z_{j}=C(N-j),\mbox{ for }j=1,...,pN,\\
Z_{j}=-C(N-j),\mbox{ for }j=pN+1,...,N.
\end{array}
\]
The constant $C$ was chosen to be $C=0.01$, but the results appear to be robust to variation in the initial conditions. Figure~\ref{fig7} demonstrates that as the proportion of the inactive oscillators increases above the AT threshold as determined by the coupling strength and the mean time delay, active oscillators synchronise but lose their oscillatory behaviour and instead approach a stable steady state, while below the threshold they continue synchronised periodic oscillations. 

\section{Discussion}\label{sec4}

In  this  paper,  we have studied aging transition in a system of globally coupled active and inactive oscillators with distributed-delay coupling. Using specific examples of uniform and gamma distributions, we have been able to analytically find boundaries of the amplitude death depending on the coupling parameters and the proportion of inactive oscillators, and we have also numerically computed characteristic eigenvalues in each scenario. For the case of uniform delay distribution, our results suggest that increasing the width of the distribution for the same mean time delay allows the system to achieve aging transition for a smaller coupling strength and a smaller proportion of inactive oscillators, and the largest proportion of inactive oscillators required for AT occurs for the discrete time delay. This highlights the fact that not only time delays can have a significant effect on aging transition, but also that the details of the delay distribution play an important role, since even for the same mean time delay, AT can occur or not depending on the width of the distribution.

In the case of the gamma distribution, provided the mean time delay is sufficiently large, there exists a range of coupling strengths, for which it is possible to achieve aging transition for any proportion of inactive oscillators, and the range of this coupling strength reduces with decreasing mean time delay. When one compares the behaviour of the system with a weak and strong distribution kernels, it becomes apparent that although AD regions exhibit qualitatively similar features for these two distributions, in the case of a strong distribution kernel aging transition occurs for higher values of the mean time delays and a higher proportion of inactive oscillators. This again reiterates the important role played by the delay distribution in quenching oscillations in coupled oscillator networks.

There are several directions in which the work presented in this paper could be extended. One interesting and practically important problem concerns the analysis of aging transition in complex networks \cite{sasai15,he2013,huang14,tanaka14}, where the coupling is not global but is rather determined by a specific network topology, while distributed delays are to be expected due to the intricate nature of connections between nodes. In this respect, understanding the dynamics of interactions between network topology and distributed-delay coupling could provide significant insights into network behaviour and robustness. Quite often it may not be practically possible to fix a specific delay distribution for the coupling \cite{barrio06,gjurchinovski2008,gjurchinovski2010}, systems may have combinations of discrete and distributed delays \cite{zou2013,rahman15}, or delays can depend on the actual state of nodes \cite{aiello92,insperger07}. Hence, another important research direction would be to analyse aging transition in systems with mixed, stochastic, and state-dependent delays. 

\bibliography{mybib}
\end{document}